\documentclass{article}

\usepackage{inputenc}
\newcommand{\cut}[1]{}
\usepackage{graphicx}
\usepackage{booktabs}
\usepackage{color}
\usepackage{amsmath}
\usepackage{amsthm}
\usepackage{amssymb}

\usepackage{natbib}
\usepackage{url}
\usepackage{wrapfig}
\usepackage{xcolor} 
\usepackage[noend, boxed, linesnumbered, algo2e]{algorithm2e}
\usepackage{algpseudocode}
\usepackage{algorithm}
\usepackage{algorithmicx}
\usepackage{algcompatible}
\usepackage{epstopdf}
\usepackage{multirow}
\usepackage{rotating}
\usepackage{relsize}
\usepackage{appendix}
\usepackage{bbm}
\usepackage{mathtools}

\usepackage{tabularx}
\usepackage{graphicx}
\usepackage{adjustbox}

\newtheorem{theorem}{Theorem}[section]

\newtheorem{proposition}[theorem]{proposition}

\newtheorem{definition}{Definition}[section]

\usepackage{appendix}
\usepackage{amsthm,amssymb}
\usepackage{tcolorbox}

\usepackage{arxiv}

\usepackage{hyperref}       
\usepackage{url}            
\usepackage{booktabs}       
\usepackage{amsfonts}       
\usepackage{nicefrac}       
\usepackage{microtype}      
\usepackage{lipsum}		
\usepackage{graphicx}
\usepackage{natbib}
\usepackage{doi}

\title{Identifying Influential Nodes in Two-mode Data Networks using Formal Concept Analysis}

\author{{Mohamed-Hamza Ibrahim} \\
	\cut{Department of Computer Science and Engineering, \\ }Universit\'{e} du Qu\'{e}bec en Outaouais, Canada \\
	\texttt{mohamed.ibrahim@polymtl.ca} \\
	\And
	{Rokia Missaoui} \\
	\cut{Department of Computer Science and Engineering, \\ }Universit\'{e} du Qu\'{e}bec en Outaouais, Canada \\
	\texttt{Rokia.missaoui@uqo.ca}
	\AND
	{Jean Vaillancourt} \\
	\cut{Department of Decision Sciences,  \\ }HEC Montreal, Canada \\
	\texttt{jean.vaillancourt@hec.ca}
}

\renewcommand{\shorttitle}{}

\begin{document}
\maketitle

\begin{abstract}
Identifying important actors (or nodes) in a two-mode network often remains a crucial challenge in mining, analyzing, and interpreting real-world networks. While traditional bipartite centrality indices are often used to recognize key nodes that influence the network information flow, they frequently produce poor results in intricate situations such as massive networks with complex local structures or a lack of complete knowledge about the network topology and certain properties. In this paper, we introduce \textit{Bi-face} ($\text{BF}$), a new bipartite centrality measurement for identifying important nodes in two-mode networks. Using the powerful mathematical formalism of Formal Concept Analysis, the $\text{BF}$ measure exploits the faces of concept intents to identify nodes that have influential bicliques connectivity and are not located in irrelevant bridges. Unlike off-the shelf centrality indices, it quantifies how a node has a cohesive-substructure influence on its neighbour nodes via bicliques while not being in network core-peripheral ones through its absence from non-influential bridges. Our experiments on several real-world and synthetic networks show the efficiency of $\text{BF}$ over existing prominent bipartite centrality measures such as betweenness, closeness, eigenvector, and vote-rank among others.
\end{abstract}


\section{Introduction}
In today's world, complex real-life systems are ubiquitous. For example, mobile phone as well as Facebook and Twitter networks facilitate to us the way we interact with one another. Airline and railway networks provide us with the most efficient modes of transportation while also highly reducing travel times. The energy and electric power networks play a significant role in supplying our domestic and industrial lives. Most of these systems frequently feature two types of data with complex substructures and can thus be represented as two-mode networks (also known as bipartite graphs or affiliation networks). Due to the complex structure of such networks, the spread of information across the network makes some nodes more important than others in certain contexts. As such, the interesting question of how to measure the relative importance of nodes in a two-mode network is often increasingly challenging in the field of complex network analysis (CNA). As it is frequently used to understand the role of nodes within a network, node centrality analysis can provide efficient answers to this question. The centrality measure ranks nodes based on how they influence or are effected by other nodes via their connection topology. Since no consensus holds on a unique definition of centrality for two-mode networks, while opening the door for the invention of new ones, various centrality measures have been proposed in the CNA literature (cf. \citep{jackson2010social,jalili2015centiserver, oldham2019consistency} for a detailed survey), each of which takes into account a distinct aspect of a central node. In the mainstream CNA research area, the bipartite centrality is frequently classified as local or global. \cut{That is, some bipartite centrality measures quantify a node's local information within the network, whereas another measure approximates the importance of such a node based on its location in the global context of the network.}

The local centrality metrics focus on the relative importance of the node in its neighbourhood within local cohesive communities. For example, the degree centrality \citep{borgatti1997network} is a basic local metric that counts the number of links that each node has. However, \textit{it frequently captures irrelevant local information about a node in practice}. Intuitively, it is assumed that only the node with the highest degree should be in the centre (because it is the most densely linked node i.e., a hub), but it does not account for the cascade effects of its neighbour nodes. Hence, it is sometimes necessary to remove nodes with high degree values because they provide no information. For example, Angelina Jolie has a high degree centrality in Facebook's network because so many people follow her; however, if you explore your friends' Facebook pages to find out what they are interested in or who among them enjoy soccer the most, Angelina Jolie becomes completely irrelevant in that network. 

The k-shell centrality \citep{kitsak2010identification} is a community-based local centrality that enhances the degree of a node in terms of its neighbourhood connections using the k-core\footnote{A k-core of a graph $G$ is a maximal connected subgraph of $G$ such that all nodes have at least $k$ neighbours.}. Thus, the higher the portions of k-cores contain a node, the more likely it is to be a hub in the cores of a network, and thus the more important it is in a network. However, \textit{k-shell frequently produces inaccurate results when the network structure has a small number of k-cores, which is prevalent in two-mode networks}. This is due to the fact that in this case, many nodes are assigned an approximately equal number of k-cores. From the perspective of the topological graph of a two-mode network, k-bicliques may be more accurate graphical components than k-cores. That is, the number of k-bicliques among a node's neighbours is counted in order to estimate its importance using the Cross k-bicliques connectivity measure, which quantifies how the node affects information propagation through the network. However, in general, \textit{its calculation requires an exponential time and space complexity and is often sensitive to the $k$ parameter}. To compute Cross k-bicliques connectivity for a given node, we must first extract all k-bicliques from the network containing this node, which is an NP-hard problem \citep{chiba1985arboricity}. Furthermore, the determination of the optimal value of $k$ may be problematic in many applications. Strictly speaking, picking a large $k$ value may result in the overstepping of all k-bicliques with $k$ less than the chosen one, leadingto an underestimation of the influence of other nodes in local cohesive communities within the network. A small $k$ value may stimulate overestimation of the importance of other neighbour nodes, generating a behaviour similar to degree centrality.

Bipartite Closeness \citep{borgatti1997network,borgatti2011analyzing} is a common type of local centrality that is based on the geodesics. It computes the reciprocal of the sum of the distances between the node and all of the other nodes in the network. Its basic form intuitively assumes that information can efficiently flow from one node to every other node via the shortest distances. The important node is therefore the independent one that is close to other nodes in the network in terms of shortest paths. Thus, at a high level, it can address the degree centrality limitation in a few cases. However, \textit{on non-spatial networks, bipartite closeness frequently produces inaccurate results \citep{rodrigues2019network}, and its values on spatial networks tend to span a rather small dynamic range from smallest to largest}. This is because most complex real-world networks may have a high average length of the shortest path as their largest distance increases exponentially in terms of the number of nodes. That is, assuming that the minimum distance is equal to one, the asymptotic ratio between the minimum and the largest distances is $O(\frac{1}{\log n})$. This frequently implies that numerous nodes, with diverse roles in the network's information flow, may have comparable closeness scores. On the contrary, most non-spatial networks feature low geodesic distances among nodes given that high geodesic distances increase logarithmically with their network size. As a result, the dynamic range of variations, as well as the network diameter, will be too small, and even slight changes in the network structure can have a significant impact on nodal closeness values.  

The bipartite Betweenness \citep{brandes2001faster,borgatti1997network,borgatti2011analyzing} is another common geodesic-based measure. To evaluate the importance of a node, it computes the number of times it exists in the bridge along the geodesic paths among the other nodes in the network. Thus, it considers other nodes' dependence on a given node, and measures its optimal flow control on information passing among nodes, whether Closeness perceives the connection efficiency or independence from potential flow control through the use of intermediary nodes (cf. \citep{brandes2016maintaining}, a detailed study differentiating between closeness and betweenness). In general, \textit{bipartite betweenness does not consider node connectivity and its calculation is frequently time-consuming}. The fundamental assumption of betweenness is that every pair of nodes exchanges information through shortest-paths with equal probability. However, this is, in many situations, not a realistic assumption since information does not necessarily take the shortest path \citep{newman2005measure} (e.g., news related to a friend might not be directly known from  another close friend but from other mutual friends). As a result, it does not provide a precise representation of the most influential nodes within these groups, but rather a fair approximation (see \citep{newman2018networks} for a more detailed explanation). Furthermore, its exact centrality computation on large or dense two-mode networks requires a time complexity of $O(n_1^3+n_2^3)$, where $n_1$ and $n_2$ are the number of the two types of nodes, respectively.  

Looking at local centrality from a different angle, bipartite percolation centrality \citep{piraveenan2013percolation} estimates a node's relative importance by counting the number of percolated paths that pass through it. The percolated path is the shortest path between two nodes in which the source node is percolated (e.g., infected) but the target node may be not. The percolation centrality fully captures the essential mechanics of contagion-mediated network spreading by associating percolation paths with weight terms that determines how much importance is given to potential percolation paths originating from given nodes. This is indeed helps percolation centrality to avoid the limitation of both betweenness and closeness, which rely solely on topological and random diffusion processes via random shortest-paths. \textit{It may, however, produce poor results when the spread of contagion has no effect on changing the node state, and it is frequently computationally expensive to calculate}. Because the percolation through a network is affected by both the level of contagion and the network structure \citep{meyers2007contact}, the spread of contagion in a complex network (CN) may not change node states in a few scenarios. From a theoretical standpoint, there is a possibility that there is no transmissibility, and in this case, the percolated contagion spreads over the edges of a complex network without changing the state of a node to either recoverable or infected, leaving them in the default state. Moreover, computing the percolation centrality in worst-case scenarios with large bipartite networks having complex local structures requires a cubic time complexity in the two types node numbers. 

Global measures, on the other hand, consider a node's prominence in the context of the entire network. Its principle emphasizes the hypothesis that a few important neighbours can weight more than a large number of unimportant ones. That is, a node is important if it is connected to other important nodes. For example, Bipartite Eigenvector centrality \citep{borgatti1997network,borgatti2011analyzing} quantifies whether a node is central based on its connections to other high-score nodes. It estimates the number of traversals of each node through indefinite-length random walks. Intuitively, this implies that the node in the network core is more accessible than the other nodes. From a conceptual standpoint, a node's eigenvector can be thought of as the global extension of its local degree centrality, in which both count walks that begin and terminate at that node. \textit{Eigenvector may include a localization transition, which frequently results in inaccurate centrality scores}. As demonstrated in \citep{martin2014localization}, eigenvector centrality has a localization transition under the common conditions of a network regime, causing the majority of the weight of the centrality to concentrate on a small number of nodes in the network. This implies that when a network structure contains many hubs, the eigenvector weights are skewed toward some few nodes: the hub node and its neighbours have the highest eigenvector values, while the remaining nodes have identical centrality values (likely close to zero)\cut{, complicating the identification of accurate important nodes}.  

In this paper, we present \textbf{Bi-face} ($\text{BF}$), a new bipartite centrality that can be used to identify key nodes in complex networks. While we focus on two-mode networks here, but in tandem with its formulation for one-mode networks which we present in \citep{ibrahim2020cross}, its general framework can be easily modified and applied to other representations of CNs, such as multidimensional and multilayer networks \citep{dickison2016multilayer}. The guiding idea of $\text{BF}$ is to use a formal concept analysis framework to bring together the centrality aspects of cohesiveness via bicliques, network flow via bridges, and influence of important neighbour nodes for the benefit of actionable node identification. Its conceptual hypothesis is based on the fact that important nodes should be found in influential bridges and overlapping bicliques with a large number of important nodes. That is, it quantifies how a node affects, and is affected by, its important neighbours via bicliques while also connecting the densely substructures of  a network through its presence in influential bridges. Thus, it differs from betweenness in that it deems influential bridges rather than all bridges. Unlike closeness and eigenvector, it can efficiently deal with the diverse topological structures of a network, without potentially having localization transition, due to this hybridization of the influential bridges and overlapping bicliques aspects. Furthermore, it leverages the powerful mathematical formulation of Formal Concept Analysis (FCA) to overcome the limitation of Cross k-bicliques connectivity. That is to say, it utilizes concept lattice related to the network to efficiently extract concepts that capture bridges and k-bicliques from the network while being insensitive to the $k$ parameter. Technically, CF2 computation is based solely on the set of these extracted concepts, which is often quite small in comparison with polynomial functions\cut{time complexity} in terms of nodes and edges.  As a result, in contrast to percolation, it is relatively quick to compute in practice.

The paper is organized in the following manner. Section~\ref{Back} recalls some basic definitions of FCA and traditional bipartite centrality measures. Section~\ref{centSect} explains our proposed Bi-face centrality for identifying key nodes of two-mode networks in further more detail. In Section~\ref{Exp} we conduct a thorough experimental study and a discussion. Finally, Section~\ref{Con} presents our conclusions\cut{ and directions for future work}.

\section{Background}\label{Back}
\begin{figure}[htbp]
\centering
\includegraphics[width=0.85\linewidth,height=2.8in]{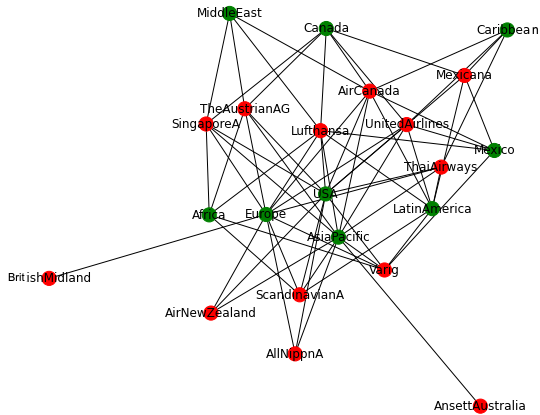}
\caption{A two-mode graph network representing flights from $13$ airline companies (in red) landing at $9$ destinations (in green) in Year 2000.}
\label{fig1}
\end{figure}

This section will briefly review the main concepts that support the comprehension of our proposed centrality measure by using an illustrative example, which is a two-mode network of airline alliances and their flying destinations in the year $2000$. As shown in Figure~\ref{fig1}, the network is modeled as an undirected bipartite graph $\Upsilon = (\mathcal{G},\mathcal{M},\mathcal{I})$, where $\mathcal{G}$ is a set of $13$ objects (also called type-I nodes) representing airline companies, $\mathcal{M}$ is a set of $9$ attributes (type-II nodes) representing flying destinations, and $\mathcal{I}$ is a set of edges where an edge $(u_i,v_j) \in \mathcal{I}$ links two nodes $u_i\in \mathcal{G}$ and $v_j \in \mathcal{M}$, if a flight from airline company $u_i$ landed at the destination $v_j$.

\subsection{Formal Concept Analysis}
In the following we recall notions of FCA \citep{Ganter+1999} that will be used in this paper.
\begin{definition}[Formal context] It is a triple $\mathbb{K} = (\mathcal{G},\mathcal{M},\mathcal{I})$, where $\mathcal{G}$ is a set of objects, $\mathcal{M}$ a set of attributes, and $\mathcal{I}$ a binary relation between $\mathcal{G}$ and $\mathcal{M}$ with $\mathcal{I} \subseteq \mathcal{G} \times \mathcal{M}$. For $g \in \mathcal{G}$ and $m \in \mathcal{M}, (g,m) \in \mathcal{I}$ holds (i.e., $(g,m)=1$) iff the object $g$ has the attribute $m$, and otherwise $(g,m) \notin \mathcal{I}$ (i.e., $(g,m)=0$). \label{defFC}
\end{definition}

Figure \ref{toycontext} is the formal context equivalent to an adjacency matrix that expresses the two-mode network shown in Figure \ref{fig1}.

Given arbitrary subsets $A \subseteq \mathcal{G}$ and $B \subseteq \mathcal{M}$, the following derivation operators are defined:
\begin{equation*}
   A^{\prime{}} = \{m \in \mathcal{M} \mid \forall g \in A, (g,m) \in \mathcal{I} \}, \; A \subseteq \mathcal{G}  
\end{equation*}
\begin{equation*}
   B^{\prime{}} = \{g \in \mathcal{G} \mid \forall m \in B, (g,m) \in \mathcal{I} \}, \; B \subseteq \mathcal{M} 
\end{equation*}
where $A^{\prime{}}$ is the set of attributes common to all objects of $A$ and $B^{\prime{}}$ is the set of objects sharing all attributes from $B$. The closure operator $(.)^{\prime{}\prime{}}$ implies the double application of $(.)^{\prime{}}$, which is extensive, idempotent and monotone. The subsets $A$ and $B$ are closed when $A=A^{\prime{}\prime{}}$, and $B=B^{\prime{}\prime{}}$. 
\begin{definition}[Formal concept]
The pair $c=(A,B)$ is called a \textit{formal concept} of $\mathbb{K}$ with \textit{extent} $A$ and \textit{intent} $B$ if both $A$ and $B$ are closed and $A^{\prime{}}=B$, and $B^{\prime{}}=A$. 
\end{definition}

The object concept $g\in \mathcal{G}$ is expressed by ${\gamma}g:=\left(g'',g'\right)$ and the attribute concept of $m\in \mathcal{M}$ is defined by ${\mu}m:=\left(m',m''\right)$. 


\begin{definition}[Partial order relation $\preceq$]
A concept $c_1=(A_1,B_1)$ $\preceq$ $c_2=(A_2,B_2)$ if:
\begin{equation}
A_1 \subseteq A_2 \iff  B_1 \supseteq B_2.    
\end{equation} 
\end{definition}
In this case, $c_2$ is called a \textit{superconcept} (or successor) of $c_1$, and $c_1$ is called a \textit{subconcept} (or predecessor) of $c_2$. The set of all concepts of the formal context $\mathbb{K}$ is expressed by $\mathcal{C}(\mathbb{K})$ or simply $\mathcal{C}$.

\begin{definition}[Concept Lattice]
The concept lattice of a formal context $\mathbb{K}$, denoted by $\mathfrak{B} (\mathbb{K})=(\mathcal{C},\preceq)$, is a Hasse diagram that represents all formal concepts $\mathcal{C}$ together with the partial order that holds between them. In $\mathfrak{B} (\mathbb{K})$, each node represents a concept with its extent and intent while the edges represent the partial order between concepts. 
\end{definition}

Figure \ref{toylattice} is the Hasse diagram of the concept lattice that corresponds to the context of Figure \ref{toycontext}. More precisely, it is a diagram with reduced labeling. This means that the label $g$ is written below ${\gamma}g$ and $m$ above ${\mu}m$. The extent of a concept represented by a node $a$ is given by all labels in $\mathcal{G}$ from the node $a$ downwards, and the intent by all labels in $\mathcal{M}$ from $a$ upwards. 

There are several methods (cf. \citep{Ganter+1999,valtchev2002partition,choi2009faster}) that build the lattice, i.e.,
compute all the concepts together with the partial order. 

\begin{definition}[Lower and Upper covers]\label{LUc}
For any two formal concepts $c_1=(A_1,B_1)$ $\preceq$ $c_2=(A_2,B_2)$ if:
\begin{align}
(A_1,B_1) \preceq (A_2,B_2), \nexists \; c3=(A_3,B_3) \text{  such that } \notag \\
(A_1,B_1) \preceq  (A_3,B_3) \preceq (A_2,B_2),   
\end{align}
or 
\begin{equation}
A_1 \subseteq A_3 \subseteq A_2 \iff  B_1 \supseteq B_3  \supseteq B_2,
\end{equation}
then $c_1=(A_1,B_1)$ is a \textit{lower cover} of $c_2=(A_2,B_2)$, and $c_2=(A_2,B_2)$ is an \textit{upper cover} of $c_1=(A_1,B_1)$; represented as $c_1 \prec c_2$ and  $c_2 \succ c_1$ respectively. 
\end{definition}
We will use $\mathcal{U}(c)$ and $\mathcal{L}(c)$ to denote the sets of upper and lower covers of the formal concept $c$ respectively.

\begin{definition}[Concept Intentional Face \citep{pfaltz2002scientific}]\label{intface} 
The \textit{intentional face} $f_{\text{in}}(c,c_d)$ of a concept $c=(A,B)$  w.r.t. its d-th upper cover concept, $c_d=(A_d,B_d) \in \mathcal{U}(c)$, is the difference between their intent sets as:
$f_{\text{in}}(c,c_d) = B \setminus B_d$.
\end{definition}
\begin{definition}[Concept Extensional Face]\label{exface} 
The \textit{extensional face} $f_{\text{ex}}(c,c_l)$ of a concept $c=(A,B)$  w.r.t. its l-th lower cover concept, $c_l=(A_l,B_l) \in \mathcal{L}(c)$, is the difference between their extent sets as:
$f_{\text{ex}}(c,c_l) = A \setminus A_l$.
\end{definition}

\begin{definition}[Blocker \citep{pfaltz2002scientific}]\label{minblo}
Given the family of faces $\Lambda_c$, the set $Z$ is said to be a \textit{blocker}  of $\Lambda_c$ if $\forall f_i \in \Lambda_c, \; f_i \cap Z \neq \emptyset$, and the blocker $Z$ is said to be \textit{minimal} if $\nexists Z_j \subset Z,\; \forall f_i \in \Lambda_c, \; f_i \cap Z_j \neq \emptyset$.
\end{definition}

\begin{definition}[Generator \citep{bastide2000mining}]\label{ming} 
Given a concept $c=(A,B)$ in a formal context $\mathbb{K} = (\mathcal{G},\mathcal{M},\mathcal{I})$, a subset $H \subseteq B$ is called a \textit{generator} of $c$ iff $H^{\prime{}\prime{}}=B$, and it is a \textit{minimal generator} when $\nexists H_1 \subseteq H$ such that $H_1^{\prime{}\prime{}} =B$.
We use $\mathcal{H}^{\text{ex}}_{c}$ and $\mathcal{H}^{\text{in}}_{c}$ to denote the sets of minimal generators of a concept $c$ w.r.t. its extent and intent respectively.
\end{definition}

\subsection{Social Network Analysis}

\begin{definition}[Biclique]\label{biclique} Let $\Upsilon = (\mathcal{G}, \mathcal{M}, \mathcal{I})$ be an undirected bipartite graph defined over the objects $\mathcal{G}$ and attributes $\mathcal{M}$. A biclique $\tilde{Q}= (\tilde{\mathcal{G}}, \tilde{\mathcal{M}})$ is a complete subgraph of $\Upsilon$  induced
by a pair of two disjoint subsets $\tilde{\mathcal{G}} \subseteq \mathcal{G}, \tilde{\mathcal{M}} \subseteq \mathcal{M}$, such that $\tilde{\mathcal{G}} \neq\emptyset$, $\tilde{\mathcal{M}} \neq\emptyset$, $\forall u \in \tilde{\mathcal{G}}$, $\forall v \in \tilde{\mathcal{M}}$, $(u,v) \in \mathcal{I}$.
\end{definition}
The disjoint subsets $\tilde{Q}=(\{\text{AirCanada}, \text{Mexicana},\text{ThaiAirways},\text{UnitedAirlines}\}$, \\ $\{\text{LatinAmerica}, \text{Caribbean}, \text{USA}\})$ is an example of a biclique. In the sequel, we use $\tilde{Q}$ as our illustrative biclique (see the lattice node indicated by a red arrow in Figure \ref{toylattice}) to support the understanding of definitions and principles related to the Bi-face centrality.

\begin{definition}[Bridge\cut{\citep{easley2010networks,granovetter1983strength}}] An edge $(u,v) \in \mathcal{I}$ of a two-mode data network $\Upsilon$ is a bridge \textit{iff} it is not contained in any cycle and its removal increases the number of connected components in the graph $\Upsilon$.
\end{definition}
For instance, the edge $(\text{AnsettAustralia},\text{AsiaPacific})$ represents a bridge in $\Upsilon$.
\begin{definition}[Bipartite centrality measure]\label{intface1} 
The \textit{centrality measure} of a type-I node $u \in \mathcal{G}$ is a function that assigns a positive real number to $u$ quantifying its centrality w.r.t. to all other type-II nodes $v \in \mathcal{M}$ in the network $\Upsilon$ (and vice versa). 
\end{definition}

The bipartite (also called two-mode) centrality measures are frequently used to identify and rank key nodes in two-mode networks. While several centrality measures have been introduced, the degree, closeness, betweenness and eigenvector have been found to be the most prominent in several applications, and they thereby are commonly used.
\begin{definition}[Degree centrality $\text{D}_{\text{c}}$ \citep{borgatti1997network,tsugawa2015analysis}]\label{Centd} The degree centrality of a node in a two-mode graph network $\Upsilon$, is defined as:
\begin{align}\label{degree}
    \text{D}_{\text{c}}(u_i) = \sum_{v_j \in \mathcal{M}} I_{ij}, \forall u_i \in \mathcal{G}, \\
    \text{D}_{\text{c}}(v_j) = \sum_{u_i \in \mathcal{G}} I_{ij}, \forall v_j \in \mathcal{M} 
\end{align}
\end{definition}
where $I_{ij}$ is equal to $1$ when a link exists between $u_i$ and $v_j$, and $0$ otherwise. Thus, the summation in Eq.~\eqref{degree}~represents the number of edges (or ties with other type neighbour nodes) involving the node.

\begin{definition}[Closeness centrality $\text{C}_{\text{c}}$ \citep{borgatti1997network,borgatti2011analyzing}]\label{Centc} The normalized closeness centrality of a node $g_i$, in a two-mode graph network $\Upsilon$, is defined as:
\begin{align}
    \text{C}_{\text{c}}(u_i) = \frac{|\mathcal{M}|+2(|\mathcal{G}|-1)}{\sum_{v_j \in \mathcal{M}} d(u_i,v_j)}, \forall u_i \in \mathcal{G},  \\
    \text{C}_{\text{c}}(v_j) = \frac{|\mathcal{G}|+2(|\mathcal{M}|-1)}{\sum_{u_i \in \mathcal{G}} d(u_i,v_j)}, \forall v_j \in \mathcal{M} 
\end{align}
\end{definition}
where $d(u_i,v_j)$  is the geodesic distance (shortest path) between the nodes $u_i$ and $v_j$.

\begin{definition}[Betweenness centrality $\text{B}_{\text{c}}$ \citep{brandes2001faster}]\label{Centb} In bipartite networks $\Upsilon$, the normalized betweenness centrality of a node is defined as in \citep{borgatti2011analyzing}:
\begin{align}
    \text{B}_{\text{c}}(u_i) = \sum_{u_j \neq u_k \neq u_i, \; u_j, u_k,u_i \in \mathcal{G}} \frac{\sigma_{u_j u_k}(u_i)}{\sigma_{u_j u_k}}, \forall u_i \in \mathcal{G},  \\
    \text{B}_{\text{c}}(v_j) = \sum_{v_j \neq v_k \neq v_i, \; v_j, v_k,v_i \in \mathcal{M}} \frac{\sigma_{v_i g_k}(v_j)}{\sigma_{v_i v_k}}, \forall v_j \in \mathcal{M},
\end{align}
where $\sigma_{x_j x_k}$ denotes the total number of shortest paths between nodes $x_j$ and $x_k$, and $\sigma_{x_j x_k}(x_i)$ is the number of those paths that traverse $g_i$. To normalize the betweenness, we simply divide $\text{B}_{\text{c}}(u_i)$ and $\text{B}_{\text{c}}(v_j)$ by the corresponding term to its node set \citep{borgatti2011analyzing}: 
\begin{equation}
    \text{B}_{\text{c}}(\mathcal{G}) = \frac{1}{2} \big[|\mathcal{M}|^2 (s+1)^2+ |\mathcal{M}| (s+1)(2t-s-1)-t(2s-t+3)\big], \forall u_i \in \mathcal{G},
\end{equation}
where $s= (|\mathcal{G}-1| \text{ div } |\mathcal{M}|)$ and $t= (|\mathcal{G}-1| \mod |\mathcal{M}|)$,  
\begin{equation}
    \text{B}_{\text{c}}(\mathcal{M}) = \frac{1}{2} \big[|\mathcal{G}|^2 (p+1)^2+ |\mathcal{G}| (p+1)(2r-p-1)-r(2p-p+3)\big], \forall v_j \in \mathcal{M},
\end{equation}
where $p= (|\mathcal{M}-1| \text{ div } |\mathcal{G}|)$ and $r= (|\mathcal{M}-1| \mod |\mathcal{G}|)$
\end{definition}

\begin{definition}[Eigenvector centrality $\text{EV}_{\text{c}}$ \citep{borgatti1997network,borgatti2011analyzing}]\label{Cente} The eigenvector centrality of a node $g_i$, in a graph network $\Upsilon$, can be iteratively computed as:
\begin{align}
    \text{EV}_{\text{c}}(u_i) = \frac{1}{\lambda} \sum_{v_j \in \mathcal{M}} a_{u_i v_j} \text{EV}_{\text{c}}(v_j), \forall u_i \in \mathcal{G}, \\
    \text{EV}_{\text{c}}(v_j) = \frac{1}{\lambda} \sum_{u_i \in \mathcal{G}} a_{u_i v_j} \text{EV}_{\text{c}}(u_i), \forall v_j \in \mathcal{M},
\end{align}
where the eigenvalue $\lambda \neq 0$ is a constant, and $a_{u_i v_j}$ is the adjacency element which is equal to $1$ if node $u_i$ is linked to node $v_j$, and $0$ otherwise.
\end{definition}

\section{Bi-face Framework}\label{centSect}
At a conceptual level, our overall Bi-face centrality approach contains the following basic steps.
\begin{enumerate}
	\item We construct the formal context associated with the network and then its corresponding concept lattice. The set of bicliques coincide with the set of formal concepts whose extent or intent is not empty.
	\item  We refine the bicliques by removing non-influential nodes in order to obtain \emph{face bicliques} (see Definition \ref{facebi}).
	\item We identify what we call \emph{face bridges}, which are the non-influential bridges in the network that contain terminal nodes
	\item We compute the Bi-face centrality measure to identify key nodes.
\end{enumerate}
 
\subsection{Building the Formal context of a Two-mode Network}
 We first build the formal context of the two-mode network $\Upsilon = (\mathcal{G},\mathcal{M},\mathcal{I})$ by computing the \textit{adjacency matrix} as follows:
\begin{equation}\label{Fcontext}
\begin{array}{lll}
\Tilde{\mathbb{K}}=(\mathcal{G},\mathcal{M},\mathcal{I}) =\begin{cases}
(u_i,v_j) = 1, & \exists~(u_i,v_j) \in \mathcal{I}, u_i \in \mathcal{G}, v_j \in \mathcal{M}\\
(u_i,v_j) = 0, & \text{Otherwise}.
 \end{cases}
 \end{array}
\end{equation}
In Eq.~\eqref{Fcontext}, we assign $1$ to the element of $\Tilde{\mathbb{K}}$ in the row $i$ and column $j$ if the object $u_i$ (node type-I) is linked to the attribute $v_j$ (node type-II) in the network $\Upsilon$. Otherwise, we assign $0$ to it. For example, the constructed formal context $\Tilde{\mathbb{K}}$ of our toy graph in Figure~\ref{fig1} is represented in \cut{Table ~\ref{toycontext}}the table of Figure~\ref{toycontext}.

\begin{figure}[htbp]
\centering
\includegraphics[width=0.97\linewidth,height=2.6in]{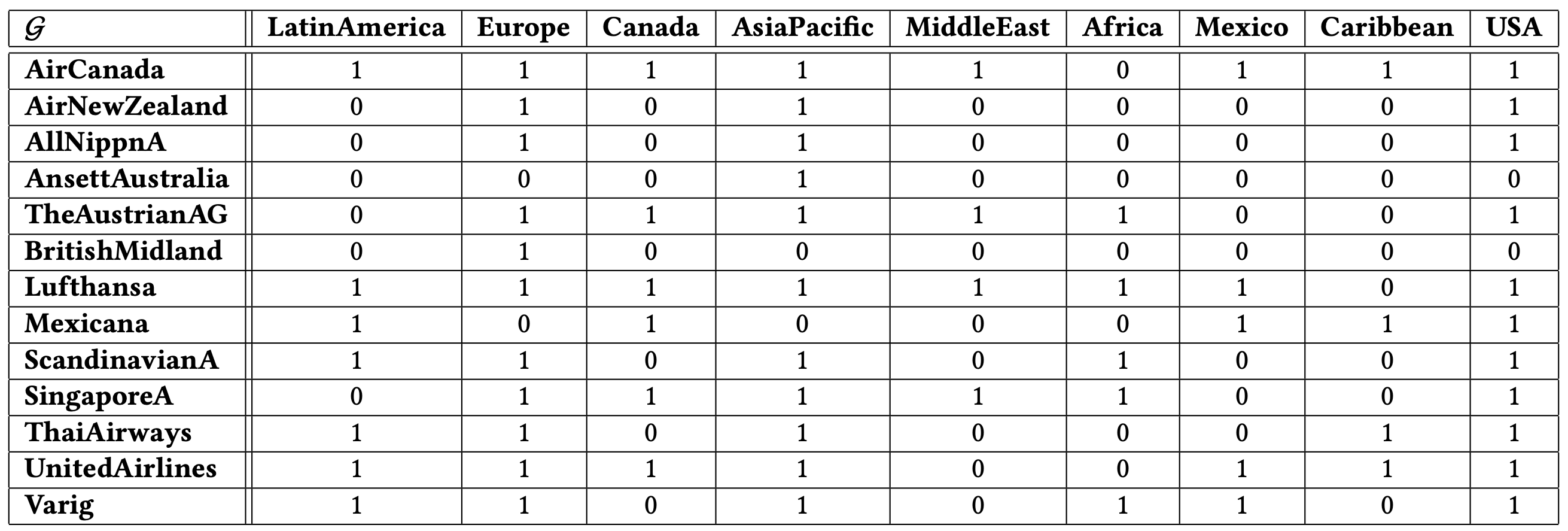}
\caption{The formal context $\Tilde{\mathbb{K}}$ for the two-mode network of Figure~\ref{fig1}.}
\label{toycontext}
\end{figure}

\begin{figure}[!htbp]
\centering
    \includegraphics[width=0.65\linewidth,height=3.3in]{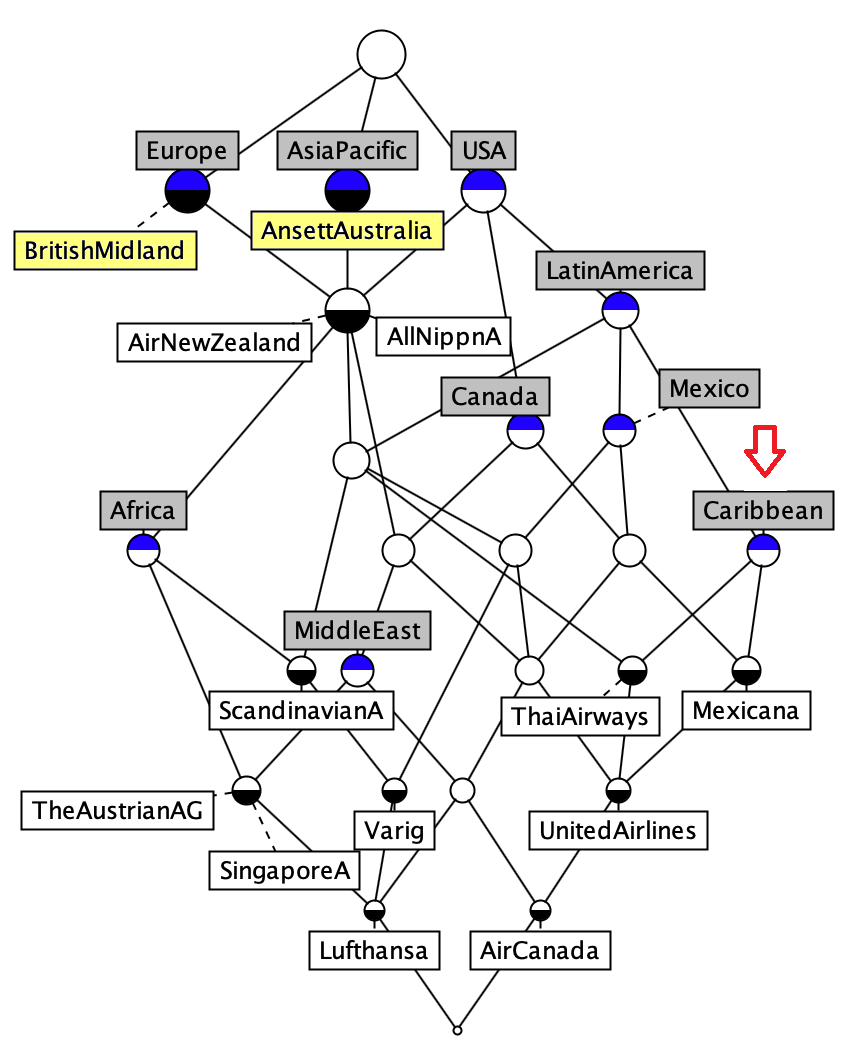}  
      \caption{The Hasse diagram of the concept lattice $\mathfrak{B}(\Tilde{\mathbb{K}})$ that corresponds to the context of the two-mode network in Figure~\ref{fig1}. More precisely, it is a diagram with reduced labeling. This means that the label $g$ is written below ${\gamma}g:=\left(g'',g'\right)$ and $m$ above ${\mu}m:=\left(m',m''\right)$. The extent of a concept represented by a node $a$ is given by all labels in $\mathcal{G}$ from the node $a$ downwards, and the intent by all labels in $\mathcal{M}$ from $a$ upwards. The red downward arrow indicates the illustrative biclique cited after Definition~\ref{biclique}.}
      \label{toylattice}
\end{figure}

\cut{
\begin{table}[!htbp]
\centering
\caption{The formal context $\Tilde{\mathbb{K}}$ for the two-mode network of Figure~\ref{fig1}.}
\begin{adjustbox}{width=\textwidth}
\begin{tabular}{|l||c|c|c|c|c|c|c|c|c|}
\hline
\textit{$\mathcal{G}$} & \textbf{\scriptsize LatinAmerica} & \textbf{ \scriptsize Europe} & \textbf{\scriptsize Canada} & \textbf{\scriptsize AsiaPacific} & \textbf{\scriptsize MiddleEast} & \textbf{\scriptsize Africa} & \textbf{\scriptsize Mexico} & \textbf{\scriptsize Caribbean} & \textbf{\scriptsize USA} \\ \hline \hline
\textbf{\scriptsize AirCanada} & 1 & 1 & 1 & 1 & 1 & 0 & 1 & 1 & 1 \\ \hline
\textbf{\scriptsize AirNewZealand} & 0 & 1 & 0 & 1 & 0 & 0 & 0 & 0 & 1 \\ \hline
\textbf{\scriptsize AllNippnA} & 0 & 1 & 0 & 1 & 0 & 0 & 0 & 0 & 1 \\ \hline
\textbf{\scriptsize AnsettAustralia} & 0 & 0 & 0 & 1 & 0 & 0 & 0 & 0 & 0 \\ \hline
\textbf{\scriptsize TheAustrianAG} & 0 & 1 & 1 & 1 & 1 & 1 & 0 & 0 & 1 \\ \hline
\textbf{\scriptsize BritishMidland} & 0 & 1 & 0 & 0 & 0 & 0 & 0 & 0 & 0 \\ \hline
\textbf{\scriptsize Lufthansa} & 1 & 1 & 1 & 1 & 1 & 1 & 1 & 0 & 1 \\ \hline
\textbf{\scriptsize Mexicana} & 1 & 0 & 1 & 0 & 0 & 0 & 1 & 1 & 1 \\\hline 
\textbf{\scriptsize ScandinavianA} & 1 & 1 & 0 & 1 & 0 & 1 & 0 & 0 & 1 \\ \hline
\textbf{\scriptsize SingaporeA} & 0 & 1 & 1 & 1 & 1 & 1 & 0 & 0 & 1 \\\hline 
\textbf{\scriptsize ThaiAirways} & 1 & 1 & 0 & 1 & 0 & 0 & 0 & 1 & 1 \\ \hline
\textbf{\scriptsize UnitedAirlines} & 1 & 1 & 1 & 1 & 0 & 0 & 1 & 1 & 1 \\ \hline
\textbf{\scriptsize Varig} & 1 & 1 & 0 & 1 & 0 & 1 & 1 & 0 & 1 \\ \hline
\end{tabular}\label{toycontext}
\end{adjustbox}
\end{table}}
We then construct the concept lattice $\mathfrak{B}(\Tilde{\mathbb{K}})$ from the formal context, as it is shown in Figure~\ref{toylattice}. Note that Figure \ref{toylattice} shows the Hasse diagram of $\mathfrak{B}(\Tilde{\mathbb{K}})$ with reduced labelling, where the label $g$ is written below ${\gamma}g$ and $m$ above ${\mu}m$. The extent of a concept represented by a node $a$ is given by all labels in $\mathcal{G}$ from the node $a$ downwards, and the intent by all labels in $\mathcal{M}$ from $a$ upwards.

\subsection{Overlapping Biclique Extraction and Refinement} 
Using the constructed lattice $\mathfrak{B}(\Tilde{\mathbb{K}})$, it is now possible to extract concepts that capture the corresponding bicliques of the two-mode network as follows:
\begin{proposition}\label{pro1} Given a network $\Upsilon$ and its corresponding concept lattice $\mathfrak{B} (\Tilde{\mathbb{K}})$, a concept $c=(A,B) \in \mathfrak{B} $ with $|A|\geq 1$ and $|B|\geq 1$, represents a biclique $Q= (\{u: u \in A \},\{v : v \in B\})$ in $\Upsilon$.
\begin{proof}
A concept represents a unit rectangular matrix of size $|A| \times |B|$ - as a sub-matrix of the adjacency matrix - and hence a biclique since it is a maximal rectangular in the formal context. Assume now that $Q= (\{u: u \in A \},\{v : v \in B\})$ is a biclique of $\Upsilon$. Then, from Definition~\ref{biclique}, for any two different nodes $u,v \in Q$, there exists an edge $(u, v)$ in $\Upsilon$ that links the two nodes. Based on Eq.~\eqref{Fcontext}, the obtained $|A| \times |B|$ adjacency matrix  $\Tilde{\mathbb{K}}(\{u: u \in A \},\{v : v \in B\},\mathcal{I}_{Q})$ that expresses the biclique $Q$ obviously represents a sub-matrix consisting of all 1's. Such a sub-matrix coincides with the concept $c=(A, B)$ in which both extent $A$ and intent $B$ involve only the objects $\{u: u \in A \}$ and attribute $\{v : v \in B\}$ nodes of $Q$ respectively. This entails that a biclique $Q$\cut{$Q= (\{u: u \in A \},\{v : v \in B\})$} is identical to a concept $c=(A,B)$.  
\end{proof}
\end{proposition}

An interesting question that could be raised now is how to determine the non-influential nodes in a given concept (or biclique). To answer this question, let us define a non-influential node from the viewpoint of FCA.
\cut{\subsection{Bicliques Refinement}}
\begin{definition}[Non-influential node]\label{influentialnode}
For a formal concept (biclique) $c_i=(A_i,B_i) \in \mathcal{C}$, a type-I node $u \in A_i$ is \textit{non-influential} if its removal from  $c_i$ (and accordingly from the graph $\mathcal{G}$) does not violate the closure conditions of other biclique concepts $\mathcal{C} \setminus \{c_i\}$ that involve it:
\begin{equation}\label{dd1}
\forall c_j \in \mathcal{C} \setminus \{c_i\} \text{ and } u \in A_j, (A_j\setminus\{u\})^{''} = A_j. \; 
\end{equation}
In a dual manner, a type-II node $v \in B_i$ is \textit{non-influential} if :
\begin{equation}\label{dd2}
\forall c_j \in \mathcal{C} \setminus \{c_i\} \text{ and } v \in B_j, (B_j\setminus\{v\})^{''} = B_j. \; 
\end{equation}
\end{definition}

That is, the subset of concepts (or bicliques) that contain either node $u$ or node $v$ still maintain their local conceptual structures even after removing $u$ from their extents or $v$ from their intents. Intuitively, this means that the node $u$ or $v$ is not important since taking it off from the graph does not affect the essential connectivity of the network (e.g., the collapsing of other concepts). In fact, Definition~\ref{influentialnode} raises another interesting question of how to determine the non-influential nodes in a given biclique. Fortunately, 
the faces of its corresponding concept, w.r.t. its upper and lower covers, can provide information as to what its non-influential nodes would be. Thus, an effective strategy here to answer this question is to contrast the corresponding concept (biclique) with its lower and upper covers through extensional and intentional faces to identify its potential non-influential type-I and type-II nodes respectively. That is, the set of faces of its concept $c_i=(A_i,B_i)$, w.r.t. its lower and upper covers, share the non-influential (type-I and type-II) nodes in its (extent and intent) respectively:

\begin{align}
\forall u \in \{\cap_{c_l\in \mathfrak{B} (c_i)} f_{\text{ex}}(c_i,c_l)\} \implies (A_j \setminus \{u\})^{''} = A_j, \notag \\ \forall c_j \in \mathcal{C} \setminus \{c_i\} \text{ and } u \in A_j. \label{con1} \\ 
\forall v \in \{\cap_{c_d\in \mathcal{U}(c_i)} f_{\text{in}}(c_i,c_d)\} \implies (B_j \setminus \{v\})^{''} = B_j, \notag \\ \forall c_j \in \mathcal{C} \setminus \{c_i\} \text{ and } v \in B_j.  \label{con2}
\end{align}

For instance, the corresponding concept of $\tilde{Q}$ has two extensional faces $f_{ex}^1= \{ThaiAirways\}$ and $f_{ex}^2=\{Mexicana\}$ respectively. The intersection of these two faces is empty, which means that there is no non-influential type-I nodes in the $\tilde{Q}$. It also has only one intensional face $f_{in}^1= \{Caribbean\}$. Thus, the intersection is also $f_{in}^1$, which entails that $Caribbean$ is a non-influential type-II node in the $\tilde{Q}$.

On the basis of Equations~\eqref{con1} and~\eqref{con2}, we can leverage the faces of concepts to define a key biclique\footnote{Note that a biclique is key when all of its nodes are influential.} as follows:    

\begin{definition}[Face Biclique]\label{facebi} Given a two-mode network $\Upsilon$ and its corresponding concept lattice $\mathfrak{B} (\Tilde{\mathbb{K}})$, a concept (representing a biclique) $c=(A,B) \in \mathfrak{B} $, is called a \textit{face biclique} if all of its (type-I and II) nodes are influential, i.e., no one of them satisfies the conditions in Equations~\eqref{con1} and~\eqref{con2}. 
\end{definition}

Based on Definition~\ref{facebi}, we can obtain the face biclique $\hat{c}=(\hat{A},\hat{B})$ by refining the original biclique $c=(A,B)$ as follows: 
\begin{equation}\label{indicator1}
\begin{array}{lll}
\hat{A}
= \begin{cases}
A \setminus \{\cap_{c_l\in \mathcal{L}(c_i)} f_{\text{ex}}(c_i,c_l)\}, &  |A| > 1 \\
A, & \text{otherwise},
 \end{cases} \\ \\
 \hat{B}
= \begin{cases}
B \setminus \{\cap_{c_d\in \mathcal{U}(c_i)} f_{\text{in}}(c_i,c_d)\}, &  |B| > 1 \\
B, & \text{otherwise}.
 \end{cases}
\end{array}
\end{equation}
In Equation~\eqref{indicator1}, we remove non-influential type-I nodes from its extent and non-influential type-II nodes from its intent. It is worth noting that when the extent or intent contains only one node, no refinement is applied because this node is influential by default. This is due to the fact that removing this node clearly violates the closure conditions in Equations~\eqref{con1} and~\eqref{con2}.

\subsection{Face-Bridge Detection}

\begin{definition}[Face-I Bridge and Terminal type-I node]\label{influentialbridgeI}
Given a 2-mode network $\Upsilon$ and its corresponding concept lattice $\mathfrak{B} (\Tilde{\mathbb{K}})$, an edge $(u,B)$ represents a non-influential (face-I) bridge containing a terminal (type-I) node $u \in \mathcal{G}$ when there is an attribute concept $c=(A,B) \in \mathfrak{B}(\Tilde{\mathbb{K}}) $ with $|B|=1$ that satisfies the following:
\begin{equation} \label{terminal1}
    u \in A \text{ and } \exists h_i \in \mathcal{H}^{\text{ex}}_{c} \text{ S.t. } h_i = u \text{ and } |h_i| = 1
\end{equation}
\end{definition}
For instance, the attribute concept $c=(\{\text{AirCanada}, \text{AirNewZealand},$ \\ $\text{AllNippnA}, \text{TheAustrianAG}, \text{BritishMidland}, \text{Lufthansa}, \text{ScandinavianA},$ \\ $\text{SingaporeA}, \text{ThaiAirways}, \text{UnitedAirlines}, \text{Varig}\},\{\text{Europe}\})$ that appears in blue/black in Figure~\ref{toylattice} has an extensional minimal generator set $\mathcal{H}^{\text{ex}}_{c} =\{\text{BritishMidland}\}$. This implies that $\textit{BritishMidland}$ (in yellow in Figure~\ref{toylattice}) is a terminal (type-I) node and the edge $(\text{BritishMidland},\text{Europe})$ represents a non-influential (face-I) bridge. Similarly, we have:
\begin{definition}[Face-II Bridge and Terminal type-II node]\label{influentialbridgeII}
Given \\ a 2-mode network $\Upsilon$ and its corresponding concept lattice $\mathfrak{B} (\Tilde{\mathbb{K}})$, an edge $(A,v)$ represents a non-influential (face-II) bridge containing a terminal type-II node $v \in \mathcal{M}$ when there is an object concept $c=(A,B) \in \mathfrak{B}(\Tilde{\mathbb{K}}) $ with $|A|=1$ that satisfies the following:
\begin{equation} \label{terminal2}
    v \in B \text{ and } \exists h_j \in \mathcal{H}^{\text{in}}_{c} \text{ S.t. } h_j = v \text{ and } |h_j|= 1
\end{equation}
\end{definition}

\begin{algorithm}[!htb]
\caption{\textit{Minigen()} procedure for computing the intentional minimal generators of a concept intent.}
\KwIn{Concept intent $B$, Set of upper covers $\mathcal{U}(c)$.}
\KwOut{Set of minimal generators $\mathcal{H}^{\text{in}}_{c}$.}
 \begin{algorithmic}[1] 
      \STATE $\mathcal{H}^{\text{in}}_{c} \leftarrow \emptyset$;
      \FOR{\text{each  $c_u=(A_u,B_u)$ in $\mathcal{U}(c)$}}
          \STATE $f_u \leftarrow B \setminus B_u$; 
          \IF{$\mathcal{H}^{\text{in}}_{c} == \emptyset$}    
              \STATE $\mathcal{H}^{\text{in}}_{c} \leftarrow \{a | \forall a \in f_u\}$;
          \ELSE
              \STATE $\text{Gen} \leftarrow \emptyset$;
              \FOR{\text{each $h_i$ in $\mathcal{H}^{\text{in}}_{c}$}}
                \IF{$h_i \cap f_u == \emptyset$} 
                   \STATE $\text{Gen} \leftarrow \text{(Gen} \cup \{h_i \cup a | \forall a \in f_u\}$);
                \ELSE
                   \STATE $\text{Gen} \leftarrow \text{(Gen} \cup \{h_i\}$);
                \ENDIF
             \ENDFOR
             \STATE $\mathcal{H}^{\text{in}}_{c} \leftarrow \text{minimal(Gen)}$;
            \ENDIF
        \ENDFOR
       \STATE $\text{\textbf{Return}} \; \mathcal{H}^{\text{in}}_{c}$;
    \end{algorithmic} \label{Genalgo}
\end{algorithm}
The question now is, how can we obtain the minimal generators of object and attribute concepts? We can efficiently compute the set of minimal generators $\mathcal{H}^{\text{in}}_{c}$ of a concept $c$ intent by applying \textit{Minigen()} procedure, which is given in Algorithm \ref{Genalgo}. It iteratively calculates the face of $c$ w.r.t. each upper cover in $\mathcal{U}(c)$ (Line 3). If the set of intentional minimal generators is empty, it then assigns the individual attributes in the first face to $\mathcal{H}_{c}$ (Lines 4-5). Otherwise, it progressively checks the intersection between the calculated face $f_u$ and each generator $h_i$ in $\mathcal{H}^{\text{in}}_{c}$ (Line 8). If the intersection with the current generator $h_i$ is empty, then $h_i$ is not in the family blocker formed by the face (Line 9). This entails that the generator $h_i$ must then be modified so that it belongs to the minimal blocker family of faces. Thus, the new minimal generators will be obtained by adding each element of the current face $f_u$ to $h_i$ (Line 10). If the intersection is not empty, then the current generator $h_i$, which exists in the family of minimal blockers of previous faces, is also a minimal blocker of the family formed of the the current face $f_u$. So, we add the generator $h_i$, without performing any modification to the minimal generator set $\mathcal{H}^{\text{in}}_{c}$ (Line 12). It ultimately verifies the minimality of the obtained set (Line 15) and returns the final set of minimal generators $\mathcal{H}^{\text{in}}_{c}$ (Line 18). Note that, in a dual way and using the set of concept's lower-covers $\mathcal{L}(c)$, we can apply $\textit{Minigen()}$ procedure to compute the set of extensional minimal generators $\mathcal{H}^{\text{ex}}_{c}$ of a concept w.r.t. its extent $A$.

\subsection{Bi-face Centrality}
\begin{definition}[Bi-face Centrality $\text{BF}_{\text{c}}$] The Bi-face centrality of nodes $ u \in \mathcal{G}$ and of $ v \in \mathcal{M}$, in a given graph network $\Upsilon$, can be computed as:
\begin{align}\label{Concent}
      \text{BF}_{\text{I}}(u) = \overbrace{\frac{|\{\hat{c} \in \Hat{\mathcal{C}} \mid u \in \hat{A} \cut{A_{\hat{c}}}|\} }{|\Hat{\mathcal{C}}|}}^{\text{Face-bicliques containing $u$}} + \big[1- \overbrace{\frac{|\{g \in \Gamma_I \mid g == u|\} }{|\Gamma_I|}}^{\text{Face-I bridges containing $u$}}\big], \\ 
      \text{BF}_{\text{II}}(v) = \overbrace{\frac{|\{\hat{c} \in \Hat{\mathcal{C}} \mid v \in \hat{B} \cut{B_{\hat{c}}}|\} }{|\Hat{\mathcal{C}}|}}^{\text{Face-bicliques containing $v$}} + \big[1-\overbrace{\frac{|\{m \in \Gamma_{II} \mid m == v|\} }{|\Gamma_{II}|}}^{\text{Face-II bridges containing $v$}}\big].\label{Concent1}
\end{align}
\end{definition}
$ \Hat{\mathcal{C}}$ stands for the set of face bicliques while $\Gamma_I$ and $\Gamma_{II}$ represent the two sets of non-influential (face-I) and (face-II) bridges, respectively. In Eq.~\ref{Concent}, the Bi-face centrality computes the sum of \textit{face-biclique}\footnote{Note that the face-clique of a node is the number of overlapping face bicliques to which it belongs to.} and \textit{Face-bridge} terms. The numerator of the face-biclique of the first term simply counts the number of refined concepts, with extent and intent sizes greater than 1, that involve a type-I node $u$. Thus, it quantifies the portion of face bicliques, in the graph network $\Upsilon$, which the node $u$ belongs to. From a conceptual perspective, this term can be considered as an efficient way of computing the cross connectivity \citep{faghani2013study,everett1998analyzing} of the node $u$ using refined overlapped bicliques that only contain influential nodes.  In the face-bridge term, we first quantify the ratio of the face bridges that involve the node $u$. This ratio is then subtracted from $1$ to approximate the portion of influential bridges in the graph that contain the node $u$. Note that the numerators of both face biclique and Face-bridge terms are unnormalized quantities. Thus, the denominators in Eq.~\ref{Concent} serve as normalization constants to scale the two terms between $0$ and $1$. In a similar manner, the Bi-face centrality in Eq.~\eqref{Concent1} can be interpreted and used to compute the centrality of type-II nodes in the graph.

Algorithm~\ref{cenalgo} gives the pseudo-code for computing the Bi-face centrality of all type-I nodes in the two-mode network $\Upsilon$. The algorithm takes as input the set of all extracted concepts $\mathcal{C}=\big\{c_j=(A_j,B_j)\big\}_{j=1}^{|\mathcal{C}|}$. For each type-I node $u_i \in \mathcal{G}$, it first iteratively refines the extents of the bicliques to obtain the face ones by removing all their non-influential type-I nodes (lines 4-5). It then counts the number of those refined face bicliques in the graph that involve $u_i$ (lines 7-9). Hereafter, it iteratively calculates the minimal generators of the the attribute concepts w.r.t. their extents to identify the face-bridges that involve the node $u_i$ (lines 11-12). It then counts the number of those face-bridges that involve the node $u_i$ as a terminal (type-I) one (lines 13-15). Subsequently, it computes the Bi-face centrality $\text{BF}_{\text{I}}$ of a node $u_i$ (lines 19-21). Finally, it returns a list containing the Bi-face centrality measures $\text{BF}_{\text{I}}$ of all type-I nodes in the graph respectively (line 22). Without loss of generality, and in a dual manner, algorithm~\ref{cenalgo} can be applied to compute the Bi-face centrality for each type-II node $v_j \in \mathcal{M}$ as follows. It iteratively obtains the face bicliques by refining the non-influential type-II nodes from the intents of their corresponding concepts. It then identifies the face bicliques in the graph that involve $v_j$. It then uses the minimal generators of object concepts to count the number of the face-bridges that involve the node $v_j$ as a terminal (type-II) one. Finally, it returns a list containing the Bi-face centrality measures $\text{BF}_{\text{II}}$ of all type-II nodes in the graph. 

\cut{Note that, based on the obtained bi-face centrality lists, we can rank the nodes in a descending order according to their importance. Table~\ref{Toycentrality} summarizes the ranked lists of the most important airlines and destinations, in Figure~\ref{fig1}, based on five bipartite centrality measures: Bi-face, betweenness, eigenvector, closeness and degree. For example, because the node $\textit{Lufthansa}$ has slightly fewer geodesics than $\textit{AirCanada}$, Betweenness considers $\textit{AirCanada}$ to be the most important type-I node. In contrast, the Bi-face centrality ranks the node $\textit{Lufthansa}$ as the most important type-I node because $\textit{Lufthansa}$ exists in considerably more overlapped bicliques than $\textit{AirCanada}$. Closeness, degree, and eigenvectors are unable to distinguish which node $\text{Lufthansa}$ or $\textit{AirCanada}$ is more important than another. Furthermore, neither degree nor closeness centrality can determine which type-I node from $\{\textit{TheAustrianAG}, \textit{SingaporeA, Varig}\}$ is more influential than the others. The eigenvector centrality cannot distinguish between type-II nodes in $\{MiddleEast, Africa, Caribbean\}$. }  

\begin{algorithm}[!htbp]
\caption{Computing Bi-face centrality ($\text{BF}_{\text{c}}$) for all type-I nodes in a two-mode network.}
\KwIn{Set of bicliques ($\mathcal{C}=\big\{(A_j,B_j)\big\}_{j=1}^{|\mathcal{C}|}$).}
\KwOut{Bi-face centrality ($\text{BF}_{\text{I}}$) of all type-I nodes.}
 \begin{algorithmic}[1]  
      \STATE $\text{BF}_{\text{I}} \leftarrow \Gamma_{I} \leftarrow \emptyset$; 
      \cut{\textit{// 1. Calculate centrality for type-I nodes} }
      \FOR {each $u_i \in \mathcal{G}$}
       \STATE $\text{count}_I \leftarrow  \gamma_{I} \leftarrow [0]_{i=1}^{|\mathcal{G}|}$;
       \FOR{$\text{each } A_j \in \mathcal{C}$}
             \STATE $\hat{A}_j \leftarrow \text{ Refine($A_j$)}; \textit{ // using Eq.} \ref{indicator1}$
             \STATE \textit{// Counting face bicliques that contain the node $u_i$}
             \IF{$|\hat{A}_j| > 0$ and $u_i \in \hat{A}_j$}
                 \STATE $\text{count}_I[i] \leftarrow \text{count}_I[i]+1$;
             \ENDIF \\ 
        \textit{// Counting face-bridges that contain the node $u_i$}
             \IF{$|B_j| == 1$} 
             \STATE $\mathcal{H}^{\text{ex}}_{A_j} \leftarrow Minigen(A_j); \textit{// using the extensional version of Algorithm~\ref{Genalgo}.}$ 
              \IF{$\exists h \in \mathcal{H}^{\text{ex}}_{c_j}, h==u_i$}
                 \STATE $\gamma_{I}[i] \leftarrow \gamma_{I}[i]+1; \Gamma_{I}.append(u_i)$;
                 \ENDIF
             \ENDIF
         \ENDFOR 
         \ENDFOR
         \FOR {each $i=1$ to $|\mathcal{G}|$}
         \STATE $\text{BF}_{\text{I}}[i] \leftarrow \big(\text{count}_I[i]/|\mathcal{C}|\big) + \big(1-(\gamma_{I}[i]/|\Gamma_{I}|)\big)$;
         \ENDFOR
       \STATE \textbf{Return}  $\text{BF}_{\text{I}}$
   \end{algorithmic} \label{cenalgo}
\end{algorithm}

\paragraph{Complexity Analysis} The calculation of the face biclique term has time and space complexity equal to $O(|\mathcal{C}|)$ since we store and proceed through the extent of all the bicliques to count the face bicliques that contain the node. The Face-bridge term of type-I node needs iterating through the attribute concepts $\tilde{\mathcal{C}}$ and calculates their minimal generators w.r.t. their corresponding lower covers. Thus, the Bi-face centrality $\text{BF}_{\text{I}}$ of all type-I nodes requires $\big(|\mathcal{G}|\times |\mathcal{C}|+|\tilde{\mathcal{C}}| \times |\tilde{\mathcal{L}}| \times |\tilde{\mathcal{H}^{\text{ex}}}|)$, where $\tilde{\mathcal{C}}$ is the set of attribute concepts, $|\tilde{\mathcal{H}^{\text{ex}}}|$ is the largest size of an obtained set of minimal generators for attribute concepts, and $\tilde{\mathcal{L}}$ is the largest number of lower covers for an attribute concept. Now, since we often have $|\tilde{\mathcal{C}}| \ll |\mathcal{C}|$ and also $|\tilde{\mathcal{L}}| \ll |\mathcal{G}|$, then the first term frequently dominates the second one. This entails that computing the Bi-face centrality $\text{BF}_{\text{I}}$ of all type-I nodes needs a time and space complexity of $O(|\mathcal{G}| \times |\mathcal{C}|)$. In a dual way, the calculation of the Bi-face centrality $\text{BF}_{\text{II}}$ of all type-II nodes has a time complexity of $O(|\mathcal{M}| \times |\mathcal{C}|)$. In total, the Bi-face centrality has time and space complexity of $O\big(|\mathcal{C}|\times (|\mathcal{G}|+|\mathcal{M}|)\big)$. 

\cut{\begin{table}[!htbp]
\caption{The ranking of all nodes in the two-mode network of Figure~\ref{fig1} based on five bipartite centrality measures: Bi-face ($\text{BF}_{\text{c}}$), betweenness ($\text{B}_{\text{c}}$), eigenvector ($\text{E}_{\text{c}}$), closeness ($\text{C}_{\text{c}}$) and degree ($\text{D}_{\text{c}}$).}
\centering
\begin{tabular}{|c|c|c|c|}
\hline
& \textit{\small Rank} & \textbf{\small Type-I} & \textbf{Type-II} \\ \hline
\multirow{10}{*}{\textbf{\small $\text{BF}_{\text{c}}$}} & \textbf{1} &  Lufthansa &  USA \\ 
& \textbf{2} &  AirCanada &  AsiaPacific, Europe \\ 
& \textbf{3} &  UnitedAirlines &  LatinAmerica \\ 
& \textbf{4} &  Varig &  Canada \\ 
& \textbf{5} &  SingaporeA,  TheAustrianAG &  Mexico \\ 
& \textbf{6} &  ScandinavianA &  Africa \\ 
&\textbf{7} &  ThaiAirways &  MiddleEast \\ 
& \textbf{8} &  Mexicana &  Caribbean \\ 
& \textbf{9} &  AirNewZealand,  AllNippnA & - \\
& \textbf{10} &  AnsettAustralia, BritishMidland & - \\ \hline
\multirow{8}{*}{\textbf{\small $\text{E}_{\text{c}}$}} & \textbf{1} &  Lufthansa, AirCanada &  USA \\ 
& \textbf{2} &  UnitedAirlines &  AsiaPacific, Europe \\ 
& \textbf{3} &  Varig &  Canada \\ 
& \textbf{4} &  SingaporeA, TheAustrianAG &  LatinAmerica \\ 
& \textbf{5} &  ThaiAirways, ScandinavianA &  Mexico \\ 
& \textbf{6} &  Mexicana &  MiddleEast, Africa, Caribbean \\ 
&\textbf{7} &  AllNippnA, AirNewZealand & - \\ 
& \textbf{8} &  AnsettAustralia, BritishMidland & - \\ \hline
\multirow{5}{*}{\textbf{\small $\text{C}_{\text{c}}$}} & \textbf{1} &  Lufthansa, AirCanada &  USA, AsiaPacific, Europe \\ 
& \textbf{2} &  UnitedAirlines &  LatinAmerica \\ 
& \textbf{3} &  Varig, SingaporeA, TheAustrianAG &  Canada \\ 
& \textbf{4} &  ThaiAirways, ScandinavianA &  Mexico \\ 
& \textbf{5} &  Mexicana, AllNippnA, AirNewZealand &  MiddleEast, Africa  \\ 
& \textbf{6} &  AnsettAustralia, BritishMidland &  Caribbean \\ \hline
\multirow{10}{*}{\textbf{\small $\text{B}_{\text{c}}$}} & \textbf{1} &   AirCanada &  AsiaPacific, Europe \\ 
& \textbf{2} &  Lufthansa &  USA \\ 
& \textbf{3} &  UnitedAirlines &  LatinAmerica \\ 
& \textbf{4} &  SingaporeA, TheAustrianAG &  Canada \\ 
& \textbf{5} &  Varig &  Mexico \\ 
& \textbf{6} &  ThaiAirways &  Africa \\ 
&\textbf{7} &  ScandinavianA &  Caribbean \\ 
& \textbf{8} &  Mexicana &  MiddleEast \\ 
& \textbf{9} &  AirNewZealand,  AllNippnA & - \\
& \textbf{10} &  AnsettAustralia, BritishMidland & - \\ \hline
\multirow{6}{*}{\textbf{\small $\text{D}_{\text{c}}$}} & \textbf{1} &  AirCanada, Lufthansa  &  Europe, AsiaPacific, USA \\ 
& \textbf{2} &  UnitedAirlines &  LatinAmerica \\ 
& \textbf{3} &  TheAustrianAG, SingaporeA, Varig &  Canada \\ 
& \textbf{4} &  Mexicana, ScandinavianA, ThaiAirways &  Africa, Mexico \\ 
& \textbf{5} &  AirNewZealand, AllNippnA &  MiddleEast, Caribbean \\ 
& \textbf{6} &  AnsettAustralia, BritishMidland & - \\ \hline
\end{tabular}\label{Toycentrality}
\end{table}}

\section{Experimental Evaluation}\label{Exp}
The goal of our experimental evaluation is to investigate the following key questions.
\begin{itemize}
    \item (\textbf{Q1}) Is the Bi-face centrality more accurate than the state-of-the-art centrality measures? 
     \item (\textbf{Q2}) Is Bi-face centrality performing fast compared to prominent centrality measures?
\cut{    \item (\textbf{Q3}) Is the Bi-face centrality approach correlated to other state-of-the-art centrality measures?}
\end{itemize}
To find robust answers, we first select the following four (real-life$^\star$ and synthetic$^\ddagger$) two-mode networks which have different configurations, and they thereby facilitate the validation of various scenarios.
\subsection{Datasets}
\begin{itemize}
\item $^\star$\textbf{Norwegian Interlocking Directorates} \citep{seierstad2011few}, which contains interlocking boards of $1542$ Norwegian director women in $373$ Norwegian public limited companies. A link represents a board membership connecting a woman as a director of a public company in Norway on August 2009. 
    \cut{\item $^\star$\textbf{OPSAHL-collaboration (arXiv cond-mat)} \citep{newman2011structure}, which contains authorship links between authors and publications in the arXiv condensed matter Section with $16726$ authors and $22015$ articles from 1995 to 1999. A link represents an authorship connecting an author and a paper. }
    \item $^\star$\textbf{PediaLanguages}\citep{morsey2012dbpedia} involves the semantic web of $316$ official languages spoken by people living in $169$ different countries. An edge connects an official language to a country if people in that country speak that language.
\item $^\star$\textbf{Southern-Women-Davis} \citep{borgatti20092,Freeman2003}, which is a two-mode social network of $18$ women reporting their participation in $14$ events (such as a meeting of a social club, a church event and a party) over a nine-month period. A woman is connected to an event if she attends that event
 \item $^\ddagger$\textbf{CoinToss}, which is a random bipartite network generated by indirect Coin-Toss model generator \citep{felde2020null}.  
\end{itemize}
A few statistics of the networks is summarized in Table~\ref{data}.\footnote{Datasets are publicly available at: 
https://toreopsahl.com/datasets/ \\
http://konect.cc/networks/opsahl-collaboration/ \\
https://networkdata.ics.uci.edu/netdata/html/davis.html}

\begin{table}[!htbp]
\centering
\caption{A brief statistics of the social networks, which includes the number $|\mathcal{G}|$ of type-I nodes, the number $|\mathcal{M}|$ of type-II nodes, the number $|\mathcal{I|}$ of edges, and the density $\Theta$ in $\%$.}
\label{data}
\begin{tabular}{|l|c|c|c|c|}
\hline
Name & \textbf{$|\mathcal{G}|$} & $|\mathcal{M}|$ &  \textbf{$|\mathcal{I}|$}  & $\Theta$
 \\ \hline
\textbf{Norwegian-Directorate} & $1542$ & $375$ & $1889$ & $0.33$ \\ \hline
\textbf{PediaLanguages} & $316$ & $169$ & $9022$ & $17$ \\ \hline
\textbf{Southern-Women-Davis} & $18$ & $14$ & $89$ & $37$ \\ \hline
\textbf{CoinToss} & $793$ & $10$ & $3310$ & $42$ \\ \hline
\end{tabular}
\end{table}

\subsection{Methodology}
Subsequently, we compared the results of our proposed \textbf{Bi-face} centrality with the following measures:
\begin{itemize}
    \item \textbf{Bipartite closeness} [Definition~\ref{Centc}], a prominent diameter-based centrality
    \item \textbf{Bipartite Betweenness} [Definition~\ref{Centb}], a state-of-the-art geodesics-based centrality
    \item \textbf{Bipartite Eigenvector}[Definition~\ref{Centd}], a state-of-the-art centrality that assesses the importance of a node based on its connections to other highly influential nodes in a network.
    \item \textbf{Vote-Rank} \citep{zhang2016identifying}, which is a well-known method for identifying decentralized spreaders. It calculates the ranking of the nodes in the bipartite graph based on a voting scheme. That is, at each turn, all nodes iteratively vote in a spreader. The node with the highest votes is elected iteratively, while decreasing the voting ability of the elected spreader' neighbours in the the next turn.
    \item \textbf{Percolation} \citep{piraveenan2013percolation}, which measures the proportion of percolated paths\footnote{We recall that the percolated path is the shortest one between two nodes in which the source node is percolated (i.e., infected).} that go through a given node. So, it quantifies the relative impact of nodes in various percolation scenarios based on their topological connectivity over time. The percolation state is commonly assigned a value between $0.0$ and $1.0$, with $0.5$ being the most common that we used in our experiment.
    \item \textbf{Bipartite Degree} [Definition~\ref{Centd}], which can serve as a good baseline for comparison.
\end{itemize}
To evaluate the lists of (type-I and type-II) nodes ranked by all the centrality measures, we need to compare them with the corresponding ranked  lists that are obtained by the real spreading process of the nodes. Thus, we applied the following traditional schema on each individual type of nodes \citep{chen2012identifying,zhao2019modeling,zhao2020ranking} to validate the performance of a tested centrality measure:

\begin{enumerate}
    \item Compute the centrality measure for all nodes, and then record the node ranking list 
    \item Use SIR model \citep{chen2012identifying} to simulate the spreading ability of the nodes. In the SIR model, every node belongs to one of three states: susceptible, infected, or recovered. At each step, we set only one node to be infected, the other nodes are susceptible nodes, and then investigate the information spreads in the network. Every infected node can infect its susceptible neighbours with spreading (also called infection) probability. Note that instead of considering the recovered state of each node, we focus on the influence within a time $t=10$ since the spreading in an early stage is found to be more important in practice. At the end of the SIR simulation process, we calculate the spreading efficiency for every node, and then record the node influence ranked list 
    \item Based on the centrality-based ranking list and the one generated by the SIR model, we record the joint score list $\mathfrak{B}  = \{(x_i,y_i)\}_{i=1}^{n}$, where $x_i$ and $y_i$ are the centrality-based and SIR-based measures of a node $g_i \in \mathcal{G}$, respectively. For any two randomly selected pairs $(x_i,y_i), (x_j,y_j) \in \mathfrak{B} $, if both $(x_i < x_j)$ and $(y_i < y_j)$ or if both $(x_i > x_j)$ and $(y_i > y_j)$,  they are said to be \textit{concordant}. If both $(x_i < x_j)$ and $(y_i > y_j)$ or if both $(x_i > x_j)$ and $(y_i < y_j)$, they are said to be \textit{discordant}. If $(x_i = x_j)$ and $(y_i = y_j)$, then the pair is neither concordant nor discordant. 
\end{enumerate}
Consequently, we calculate the following Kendall's tau rank correlation coefficient $\tau$ metric:
\begin{equation}\label{taucorr0}
    \tau =  \frac{2 (n_c - n_d)}{n(n-1)},  
\end{equation}
where $n_c$ and $n_d$ are the number of concordant and
discordant pairs in $\mathfrak{B} $, respectively. A high $\tau$ value indicates that the centrality measure could produce an accurate ranked list. The ideal case is when $\tau=1$ where the ranked list generated by the centrality measure is symmetrical to the ranked list generated by the real spreading process. To evaluate the accuracy of the results, we now calculate the average Kendall's tau rank correlation coefficient as follows:
\begin{equation}\label{taucorr}
    \hat{\tau} =  \frac{\tau_{I}+\tau_{II}}{2},  
\end{equation}
where $\tau_{I}$ and $\tau_{II}$ are the Kendall's tau correlation coefficients calculated using Eq.~\eqref{taucorr0} for type-I and type-II of nodes, respectively. 

To assess the scalability, we consider the average elapsed time metric as:
\begin{equation}\label{Timeeq}
    \xi =  \frac{1}{2}\big[\frac{\sum_{u_i \in  \mathcal{G}} t_i}{n} + \frac{\sum_{v_j \in  \mathcal{M}} t_j}{m}\big]
\end{equation}
where $t_{i}$ and $t_{j}$ are the elapsed times for computing the underlying centrality measure of a type-I node $u_i \in \mathcal{G}$ and a type-II one $v_j \in \mathcal{M}$, respectively. 
 
All the experiments were run on an Intel(R) Core-i7 CPU @2.6GHz computer with 16 GB of memory under MacOS Mojave. We implemented all the considered indices as an extension to NetworkX Python package. To extract formal concepts we make use of the \emph{Concepts 0.7.11} Python package, which is implemented by Sebastian Bank\footnote{Publicly available: \url{https://pypi.python.org/pypi/concepts}}. 

\subsection{Results}\label{exp}
\subsubsection{Experiment I.}\label{expi}
This experiment is devoted to answering Question 1. Each infected node has a spreading probability $\beta$ of infecting its susceptible neighbours in the SIR model simulation. As a result, and in accordance with the scheme described above, we iteratively increase the spreading probability in the range $\beta=(0,0.1]$ with increments of $0.01$. At each step-size, we compute the joint list $\mathfrak{B} $ of each centrality measure and the real spreading of the nodes for each individual type of nodes separately. We then calculate the corresponding evaluation metric $\hat{\tau}$ in Eq.~\eqref{taucorr}. 

\begin{figure}[!htbp]
  \begin{center}
   \includegraphics[width=65mm]{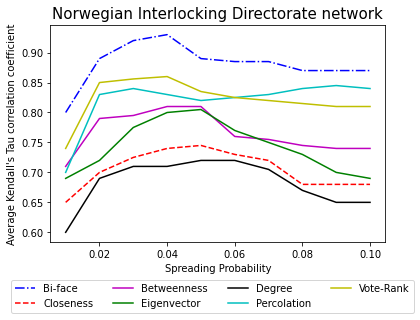} 
   \includegraphics[width=65mm]{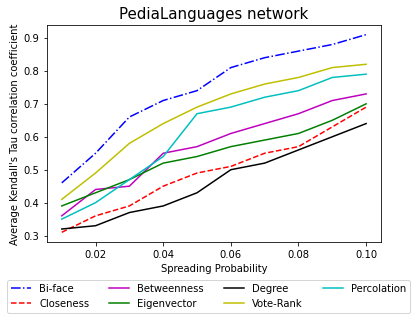} 
   \includegraphics[width=65mm]{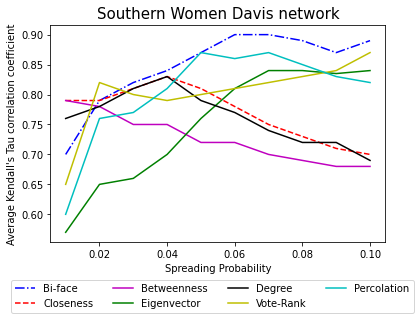}
    \includegraphics[width=65mm]{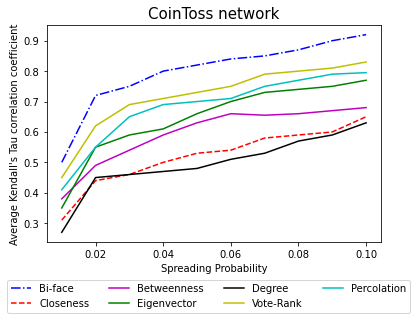}  
    \end{center}
      \caption{The average Kendall's tau coefficient $\hat{\tau}$ between the tested centrality measures and the ranking list generated by the SIR model, with $\beta \in (0,0.1]$, at $t= 10$ on the four underlying datasets.}
      \label{Exp1}
\end{figure}

Figure~\ref{Exp1} displays the average Kendall's tau correlation coefficient $\hat{\tau}$ between the seven tested centrality measures and the ranking list generated by the SIR model, with a spreading probability $\beta \in (0,0.1]$ and at a given time $t= 10$. Overall, Bi-face outperforms all the compared centrality measures, achieving the most accurate Kendall coefficient $\hat{\tau}$ on Norwegian-Directorate, PediaLanguages and CoinToss networks. On the Women-Davis network, Bi-face has the highest $\hat{\tau}$ value when the spreading probability $\beta \geq 0.03$, otherwise vote-rank, closeness, betweenness and degree slightly compete with Bi-face. The percolation comes close behind Bi-face on Women-Davis, but considerably further behind on Norwegian-Directorate, PediaLanguages and CoinToss networks. Except on the Women-Davis network with spreading probability $\beta < 0.03$, the vote-rank is clearly less accurate than Bi-face on all the tested networks, but it is more accurate than percolation, betweenness, closeness, eigenvector and degree on PediaLanguages and CoinToss networks. On the Norwegian-Directorate and Women-Davis networks, the vote-rank and percolation compete with each other. The percolation is clearly more accurate than betweenness and eigenvector when the spreading probability $\beta \geq 0.05$ on all the tested networks. Both betweenness and eigenvector dominate degree and closeness on Norwegian-Directorate, PediaLanguages and CoinToss networks. The betweenness is more accurate than eigenvector on PediaLanguages network when the spreading probability $\beta \geq 0.05$, but it is outperformed by eigenvector on CoinToss network, and both of them compete each other on Women-Davis and Norwegian-Directorate networks.

\subsubsection{Experiment II.}\label{expii}
\begin{figure}[!htbp]
  \begin{center}
   \includegraphics[width=65mm]{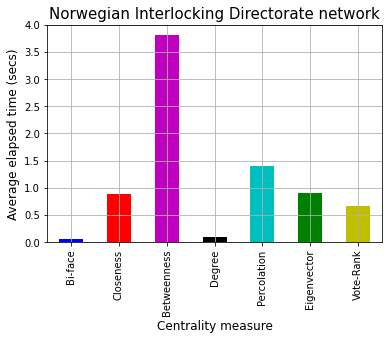} 
   \includegraphics[width=68mm]{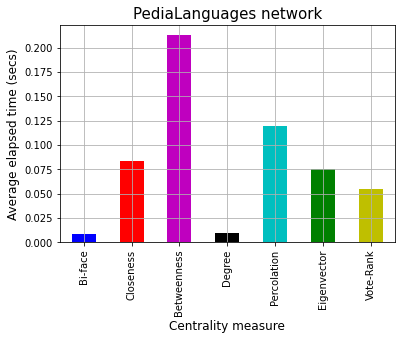} \\
   \includegraphics[width=65mm]{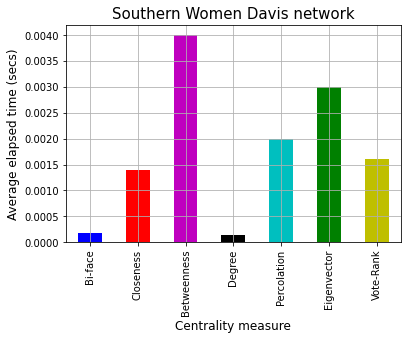}
    \includegraphics[width=65mm]{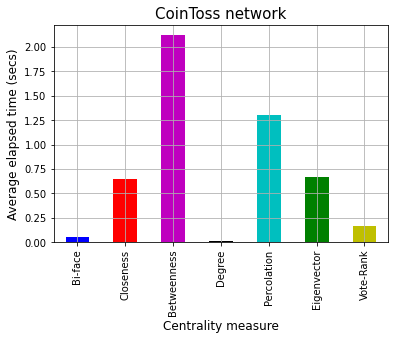}  
    \end{center}
      \caption{Average elapsed time $\xi$ (in secs) of the seven tested centrality measures: Bi-face, closeness, betweenness, degree, eigenvector, percolation and vote-rank on the four underlying datasets.}
      \label{Exp2}
\end{figure}

The second experiment is dedicated to answer Question 2. The goal here is to evaluate the performance of the centrality measures. That is, we rerun Experiment I while reporting their computational time as in Eq.~\ref{Timeeq}. The average elapsed time $\xi$ of the seven centrality measures on the four underlying networks is depicted in Figure~\ref{Exp2}. On all the tested networks, the Bi-face dominates all centrality measures (except degree). It finishes at least twenty-three times faster than betweenness, eleven times faster than percolation, nine times faster than eigenvector and eight times faster than closeness. Degree is very competitive with Bi-face on Women-Davis and CoinToss, but Bi-face clearly prevailed over the degree by a significant margin on Norwegian-Directorate and PediaLanguages networks. Apart from Bi-face, the percolation is marginally faster than both the closeness and vote-rank by at least factors of $1.3$ and $1.2$ on all networks respectively. In addition, the closeness is considerably faster than betweenness, and competes with eigenvector on Norwegian-Directorate and CoinToss networks. Vote-rank is significantly faster than closeness on Norwegian-Directorate, PediaLanguages and CoinToss networks, but on the contrary, closeness is slightly quicker than it on Women-Davis network.

\cut{\subsubsection{Experiment III.}\label{expiii}
In this experiment, we focus on Question 3. That is, we are interested here in exploring the monotonic relationships between Bi-face and the other underlying centrality measures. Table~\ref{correlation_cent} records the average Kendall's tau rank correlation coefficient between Bi-face and the other six bipartite centrality measures. Overall, all the centrality measures are positively correlated with Bi-face, which is remarkably consistent and supplement the finding of Experiment I. The Bi-face has moderate monotonic relationships with vote-rank and percolation on all tested networks. There is clearly a weak relationship between Bi-face and betweenness on the Norwegian-Directorate, PediaLanguages, and CoinToss networks. Furthermore, it has a weak correlation with eigenvector on the Women-Davis and CoinToss networks. Moreover, there are weak correlations between Bi-face and both closeness and degree on the Women-Davis network.

\begin{table}[!htbp]
\caption{Average Kendall's tau rank correlation coefficient between Bi-face ($\text{BF}_{\text{c}}$) and the other six bipartite centrality measures: betweenness ($\text{B}_{\text{c}}$), eigenvector ($\text{E}_{\text{c}}$), closeness ($\text{C}_{\text{c}}$), degree ($\text{D}_{\text{c}}$), Percolation ($\text{PC}_{\text{c}}$) and Vote-rank ($\text{VR}_{\text{c}}$) on the four underlying datasets. The moderate, weak, highly weak correlation values are represented in {\color{blue}blue}, {\color{red}red} and black respectively.}
\begin{adjustbox}{width=\textwidth}
\begin{tabular}{|c|c|c|c|c|}
\hline
 \multirow{2}{*}{} & \multicolumn{4}{c|}{\textbf{\footnotesize $\text{BF}_{\text{c}}$}} \\ \cline{2-5}
 & \textbf{\footnotesize Norwegian-Directorate} & \textbf{\footnotesize PediaLanguages} & \textbf{\footnotesize Women-Davis} & \textbf{\footnotesize CoinToss} \\ \hline
\textbf{\footnotesize $\text{E}_{\text{c}}$} & 0.09 & 0.12 & {\color{red}0.19} & {\color{red}0.20} \\ \hline
\textbf{\footnotesize $\text{C}_{\text{c}}$} & 0.07 & 0.08  & {\color{red}0.18} & 0.10 \\ \hline
\textbf{\footnotesize $\text{B}_{\text{c}}$} & {\color{red}0.14} & {\color{red}0.16} & 0.05 & {\color{red}0.19} \\ \hline
\textbf{\footnotesize $\text{D}_{\text{c}}$} & 0.04 & 0.06 & {\color{red}0.17} & 0.07 \\ \hline
\textbf{\footnotesize $\text{PC}_{\text{c}}$} & {\color{blue}0.31} & {\color{red}0.29} & {\color{blue}0.41} & {\color{blue}0.33} \\ \hline
\textbf{\footnotesize $\text{VR}_{\text{c}}$} & {\color{blue}0.32} & {\color{blue}0.33} & {\color{blue}0.39} & {\color{blue}0.35} \\ \hline
\end{tabular}\label{correlation_cent}
\end{adjustbox}
\end{table}}

\subsection{Discussion}
Taking the identification of accurate node centrality into consideration, the results of Experiment I in Subsection~\ref{expi} indicate that Bi-face outperforms traditional bipartite centrality measures such as vote-rank, percolation, degree, closeness, betweenness, and eigenvector. This is attributed to the use of its face biclique and face-bridge terms in tandem to leverage local and global aspects of network topology, respectively. That is, the face-biclique term quantifies the structural embeddedness of cohesive regions in a network involving each individual (type-I and type-II) node. From a conceptual perspective, this term considers the local information on how the node influences its immediate important neighbour nodes through the lens of its overlapping face bicliques. The face-bridge term quantifies a node's global role based on how the information flows through influential (face) bridges (i.e., important geodesics). 

In terms of effective performance, the results of Experiment II from the previous Subsection~\ref{expii}, suggest that the Bi-face is considerably faster than all other tested bipartite centrality measures (except degree). In practice, this is because Bi-face primarily calculates the centrality of all nodes based on the set of concepts $\mathcal{C}$, which is frequently too small in comparison to all other tested centrality measures with polynomial time complexity in terms of nodes and edges, i.e., $|\mathcal{C}| \ll n^p$ and $|\mathcal{C}| \ll m^q$, with $p,q >1$.

Besides that, several well-known observations are clearly consistent with the obtained results in Subsection~\ref{exp}. First, in some real-world applications, we may end up with several nodes having approximately equal low or high degrees, and in these cases, degree centrality cannot serve as a descriptive measure that can distinguish between nodes. Second, closeness can address the degree centrality limitation in a few situations. For example, consider node $u$ that is linked to node $v$. Assume that node $v$ is in close proximity to the other nodes in the network, resulting in a high closeness score. Node $u$ has a very low degree score of $1$, but a rationally high closeness score, because node $u$ can propagate information to all other nodes that node $v$ reaches with one extra step. However, closeness, like degree, is usually inappropriate for irregularly connected bipartite networks. Because the shortest-path distance between two nodes is infinite when they are not reachable through a path, the closeness score is equal (or very close) to zero for those nodes in the network that do not reach all other nodes. Third, since betweenness lacks any form of measuring local nodal connectivity, it is expected to produce relevant results only if the goal is only to quantify influence on communication among local groups, which is not always the case when studying the centrality in real-world networks. Finally, and in practice, using the efficient implementation adopted from the fastest algorithm proposed in \citep{brandes2001faster}, the calculation of percolation centrality for all nodes requires a time complexity of $O(m^2(n_1+n_2))$, which still seems to impose a computational bottleneck even with fairly medium-sized networks. 

\cut{As frequently asked, are these centrality measures correlated? The results of Experiment III in Subsection~\ref{expiii} expound that Bi-face centrality gives unique node identification based on network topology. The presence of terminal nodes, influential (also known as face) bridges, and overlapping key bicliques impacts both the performance and behaviour of Bi-face as well as its relationship to other traditional centrality measures. When the network contains a large number of cohesive regions with many nodes having high degrees and there is a small number of hole structures or terminal nodes, the role of the face-biclique term dominates the face-bridge one, and here it is anticipated that the Bi-face centrality could be partially correlated with vote-rank, degree, eigenvector and (may be) closeness centrality measures. This is due to the fact that in this scenario, the network tends to decompose into multiple bi-clusters (or two-mode communities), with the nodes with the highest degree potentially serving as the central nodes. On the flip side of the coin, when the network contains a small number of cohesive regions or a large number of sparse ones, as well as a large number of terminal nodes and bridges, the role of the face-bridge term dominates the face-biclique one, even when structural holes are present. This is due to the effect of face-bridges in determining the central nodes, and here the Bi-face centrality may be slightly correlated with percolation and betweenness. 

It is worth noting that the existence of the two scenarios, mentioned above, in the network could potentially increase Bi-face centrality to behave slightly similar to vote-rank or percolation. In an extreme scenario, such as the Women-Davis network with a large number of overlapping bicliques and no terminal nodes, the likelihood of having face-bridges decreases dramatically. This indeed imposes a harsh situation on Bi-face because it will depend solely on its face biclique term, and here it is clearly expected that Bi-face will behave similarly to degree, closeness, eigenvector, and vote-rank, but not similarly to betweenness. From a statistical perspective, the low and moderate (i.e., not high) correlations between Bi-face and other centrality measures suggest that it is, in fact, a distinct measure that is likely to be associated with different outcomes than other centrality measures. This is due to the fact that if the measures are highly correlated, they may be somewhat redundant and behave similarly.

Furthermore, one conjecture inferred from experiment results (I-III) is that two-mode network properties (e.g., density, reciprocity, centralization) may affect the correlation among bipartite centrality measures, as well as their accuracy and performance. For instance, one observation from Table~\ref{correlation_cent} and Figure~\ref{Exp1} is that as network density increases, the correlation between Bi-face and closeness, eigenvector, and degree increases, while its correlation with betweenness decreases. This observation, however, does not clearly reflect the correlation between Bi-face and both percolation and vote-rank because Women-Davis has a lower density than CoinToss, and Bi-face is more correlated with the two centrality measures on Women-Davis than on CoinToss. While this demonstrates that network density influences how well different centrality measures correlate with one another, it also indicates that the network density is not the only factor and that other network properties may have an impact. Since the study of the network properties is outside the scope of this paper, we could explore the effects of reciprocity and centralization on Bi-face in our future work.}
\section{Conclusion}\label{Con}
The detection of influential nodes in a two-mode network is frequently an important task in scientific and industrial data analysis pipelines for explaining various behaviours and outcomes. Our work here addressed an obvious gap in the present CNA literature, namely the efficient identification of key nodes by combining both local cohesiveness and global network flow aspects of centrality through the use of FCA mathematical formalization. On this basis, we devised \textit{Bi-face}, a new bipartite centrality measure that quantifies the prominence of a node in a two-mode network based on its presence in influential overlapping bicliques and bridges. While we focused on two-mode networks here, the approach can easily be modified to accommodate other complex network representations like multilayer networks. 

From a conceptual perspective, the Bi-face score is a distinct centrality in the following three elements: (i) it uses the concept lattice formulation to efficiently extract overlapping bicliques and bridges, (ii) it leverages concept faces to refine bicliques from non-influential nodes and detect influential bridges, and (iii) it exploits the fact that influential bridges and overlapping bicliques with a large number of important neighbour nodes are likely to contain key central nodes. As a result, it measures how a node affects and is influenced by its important neighbours through refined bicliques, while also linking the network dense substructures via its existence in influential bridges. According to a thorough empirical study on several synthetic and real-life two-mode networks (see Section \ref{Exp}), the Bi-face score can identify key nodes more accurately and efficiently than other state-of-the-art centrality indices such as degree, betweenness, closeness, eigenvector, percolation, and vote-rank.

\cut{There are still a few issues to consider in the future. We intend to design an online $\text{BF}$ technique that can be used to identify important nodes in dynamic complex networks, \textit{i.e.} those that evolve over time. We also plan to develop a highly fast $\text{BF}$ algorithm by parallelizing the computation of Equations~\eqref{Concent}-\eqref{Concent1}, and then explore its performance on very massive and more complicated data networks. The $\text{BF}$ formula could also be generalized to multilayer and multidimensional networks. Finally, because the $\text{BF}$ preprocessing step includes an efficient extraction of bicliques and bridges, we could modify the $\text{BF}$ approach so that it can be easily applied to other applications, such as detecting communities in a complex network and finding maximum bicliques and bridges.}

\bibliographystyle{unsrtnat}
\bibliography{References}  

\end{document}